\documentclass[prl,
 reprint,
superscriptaddress,
 amsmath,amssymb,
 aps,
]{revtex4-1}

\usepackage{graphicx}
\usepackage{dcolumn}
\usepackage{bm}
\usepackage{hyperref}

\newcommand{\Bv}{\mathbf{B}}
\newcommand{\uv}{\mathbf{u}}

\newcommand{\rey}{{\rm Re}}
\newcommand{\rem}{{\rm Rm}}
\renewcommand{\pra}{{\rm Pm}}
\newcommand{\fv}{\mathbf{f}}
\newcommand{\xv}{\mathbf{x}}
\newcommand{\vv}{\mathbf{v}}
\newcommand{\Ev}{\mathbf{E}}
\newcommand{\p}{\partial}

\newcommand{\gkyl}{{\tt Gkeyll}}

\begin{document}


\title{Dynamo in Weakly Collisional Nonmagnetized Plasmas Impeded by Landau Damping of Magnetic Fields}

\author{Istv\'{a}n~Pusztai}
\email{pusztai@chalmers.se}
\affiliation{Department of Physics, Chalmers University of Technology,
  SE-41296 Gothenburg, Sweden}
\author{James~Juno}%
\affiliation{Institute for Research in Electronics and Applied Physics,
  University of Maryland, College Park, Maryland 20742, USA}%
\author{Axel~Brandenburg}%
\affiliation{{\sc Nordita}, KTH Royal Institute of Technology and
  Stockholm University,  SE-10691 Stockholm, Sweden}%
\author{Jason~M.~TenBarge}%
\affiliation{Department of Astrophysical Sciences, Princeton University,
Princeton, NJ 08543, USA}%
\affiliation{Princeton Plasma Physics Laboratory, Princeton, New Jersey 08544, USA}%
\author{Ammar~Hakim}
\affiliation{Princeton Plasma Physics Laboratory, Princeton, New Jersey 08543, USA}%
\author{Manaure~Francisquez}%
\affiliation{Plasma Science and Fusion Center, Massachusetts Institute of Technology, Cambridge, Massachusetts, 02139, USA}%
\author{Andr\'{e}as~Sundstr\"{o}m}%
\affiliation{Department of Physics, Chalmers University of Technology,
  SE-41296 Gothenburg, Sweden}%

\date{\today}

\begin{abstract}
We perform fully kinetic simulations of flows known to produce dynamo in magnetohydrodynamics (MHD), considering scenarios with low Reynolds number and high magnetic Prandtl number, relevant for galaxy cluster scale fluctuation dynamos. We find that Landau damping on the electrons leads to a rapid decay of magnetic perturbations, impeding the dynamo. This collisionless damping process operates on spatial scales where electrons are nonmagnetized, reducing the range of scales where the magnetic field grows in high magnetic Prandtl number fluctuation dynamos. When electrons are not magnetized down to the resistive scale, the magnetic energy spectrum is expected to be limited by the scale corresponding to magnetic Landau damping or, if smaller, the electron gyroradius scale, instead of the resistive scale.  In simulations we thus observe decaying magnetic fields where resistive MHD would predict a dynamo.

\end{abstract}

\pacs{Valid PACS appear here}
\maketitle

The energy density corresponding to the microgauss ($\rm 10^{-10}\,\rm T$) magnetic field permeating the Universe at galaxy \cite{Beck2015} and galaxy cluster  \cite{carilli} scales is comparable to that of the turbulent flows \cite{Hitomi2016} on these scales.  This approximate equipartition of magnetic and directed kinetic energies is consistent with the field being generated and maintained by a turbulent dynamo (see Ref.~\cite{Schekochihin_2004} and references therein). Small seed fields are amplified by the dynamo until they become dynamically significant, after which the field strength nonlinearly saturates in a self-consistent turbulent state. Because of the multiscale and inherently three-dimensional  \cite{Cowling1933,Busse2007} nature of dynamos, they have almost exclusively been studied within the framework of magnetohydrodynamics (MHD). Although MHD is well justified for the modeling of dynamos in dense and collisional stellar interiors, it breaks down when the mean free path of the plasma particles becomes comparable with the scales of interest, such as in galaxy clusters.

Recent efforts have started to shed light on turbulent dynamos in the collisionless regime. Using kinetic tools for the ion dynamics and isothermal fluid models for the electrons, dynamo amplification of magnetic fields has been demonstrated
 \cite{Rincon3950}.
The role of pressure anisotropy instabilities, such as firehose and mirror instabilities, has been shown to be critical for dynamo amplification  \cite{StOnge}, leading to the development of sharp magnetic field line features, thereby breaking magnetic moment conservation and alleviating the issue of related stringent constraints \cite{helander_strumik_schekochihin_2016} on field growth. While the role of kinetic ions in the context of the dynamo is only just beginning to be explored, what effects, if any, kinetic electrons have on the dynamo have yet to be studied.

In this Letter, we consider a kinetic electron effect on dynamos: the Landau damping of magnetic fluctuations. This enhances the decay of magnetic perturbations compared to resistive diffusion, thereby reducing the range of scales where field amplification occurs. We also show that this effect impedes dynamo field amplification in fully kinetic simulations of weakly collisional nonmagnetized hydrogen plasmas. 
 The possibility of Landau damping of magnetic fields has not received wide attention in the literature, except for a few sporadic applications, affecting,  e.g., the persistence of magnetic fluctuations downstream of ultrarelativistic pair-plasma shock waves with consequences on synchrotron emission in gamma-ray bursts  \cite{Chang_2008}.

The turbulent dynamo is a multiscale problem: Kinetic energy injected into flows at the outer scale $l_0$, nonlinearly cascades down to viscous scales $l_\nu \sim \rey^{-3/4}l_0$, where the energy is dissipated. The scale separation, $l_0/l_\nu$, is characterized by the fluid Reynolds number, $\rey=u_0 l_0/\nu$, where $\nu$ is the kinematic viscosity and $u_0$ is the characteristic flow velocity at scale $l_0$. The dissipation scale of magnetic fluctuations, below which resistive diffusion of the fields dominates, is the resistive scale $l_\eta$. A key dimensionless quantity in dynamo theory is the magnetic Reynolds number, $\rem=u_0 l_0/\eta$, as dynamo field amplification requires a minimum $\rem$ that depends on the properties of the flow. Here $\eta=(\sigma \mu_0)^{-1}$ is the magnetic diffusivity, with the Spitzer conductivity $\sigma$, and the magnetic permeability $\mu_0$. When the magnetic Prandtl number, $\pra=\rem/\rey=\nu/\eta$ is large, as in galaxies, galaxy clusters, the intracluster medium, and in some hot accretion disks  \cite{BrandenburgSubramanian}, then $l_\eta\ll l_\nu$, and magnetic field growth mostly takes place in the range between the $l_\nu$ and $l_\eta$ scales \cite{Schekochihin_2004}. In astrophysical systems of interest, $\pra$ can be extremely large.

The physics is kinetic for scales comparable to or smaller than the Coulomb mean free path $\lambda$. Using $l_\nu \sim \rey^{-3/4}l_0$ and $\lambda\sim  \rey^{-1} l_0 M_0$, where  $M_0$ is the Mach number corresponding to $u_0$, for a moderate $\rey$ and a $M_0\sim 1$, we see that $\lambda$ and $l_\nu$ are comparable. Therefore, the scales of interest for $\pra \gg 1 $ are kinetic. 

Several processes have been proposed to generate the seed field for dynamos (see  Ref.~\cite{Subramanian19} and references therein). One of the leading candidates is the Biermann battery \cite{Biermann} at ionization fronts in the early Universe, thought to produce a typical seed field of $B\sim 10^{-24}\,\rm T$  \cite{Subramanian94,Kulsrud_1997,Gnedin_2000}.    
While galaxy clusters are magnetized down to the resistive scale at current magnetic field levels, at the time when the field was comparable to that of seed fields, the electron Larmor radius was comparable to the mean free path. That is, electrons were not magnetized on kinetic scales for the Biermann seed case, allowing for magnetic perturbations to be Landau damped.         

We consider here fully kinetic simulations of spatially periodic flows, which are known to produce a dynamo in MHD simulations. As has been done in several dynamo studies \cite{Kinney,Schekochihin_2004,StOnge},  we sacrifice the fluid cascade for numerical feasibility and focus on subviscous scales. Accordingly, $l_0$ and $l_\nu$ are comparable to our simulation box size $L_0$. The simulations employ the kinetic-Maxwell solver \citep{Juno2018} of the \gkyl{ }\citep{gkeyllsite} plasma physics simulation framework, which applies a discontinuous Galerkin method to solve the kinetic equation 
\begin{equation}
\p_t f_a +\vv \cdot \nabla f_a + \mathbf{a}_a\cdot \nabla_\vv f_a=C[f_a],
\label{kinetic}
\end{equation}
for all species $a$, with mass $m_a$, charge $e_a$, and distribution function $f_a$. In the acceleration term, $\mathbf{a}_a=\fv_a/m_a+(e_a/m_a)(\Ev+\vv\times \Bv)$, the electric and magnetic fields, $\Ev$ and $\Bv$, are computed from Maxwell's inductive equations, and $\fv_a(\xv,t)$ is an externally prescribed forcing. Inter- and intraspecies Coulomb collisions
 are modeled by a conservative Dougherty (or Lenard-Bernstein) operator  \citep{hakim2019conservative,dougherty}, $C[f_a]$. The simulations are initialized with Maxwellian electrons ($e$) and protons ($i$), with temperature $T_a=1\, \rm keV$, density $n_a=2.3\times 10^{28}\,\rm m^{-3}$, and a flow with a characteristic speed $u_0=M_0	 \sqrt{T_e/m_i}$ and $M_0=0.35$. Our baseline plasma parameters are not representative of astrophysical plasmas, rather they are chosen to give estimated values of $\rem\approx 13$ (with Spitzer resistivity) and $\rey \approx 0.64$ (with nonmagnetized collisional viscosity), thus $\pra\approx 20$ for a box size of $L_0=9.73\,\rm \mu m$ and an assumed Coulomb logarithm of $10$. The collisional mean free path is $\lambda=1.25 \,\rm \mu m$.

First, we consider the time-dependent Galloway-Proctor (GP) flow \cite{Galloway1992} that produces a fast dynamo  ($\rem$-independent growth rate, for $\rem\gg 1$) and requires a low critical $\rem$,
\begin{align}
\uv_{\rm GP} (\xv,t) = & u_0 \{
\sin(k_0 z + \sin \omega t) + \cos(k_0 y + \cos \omega t),  \nonumber \\
 &\cos(k_0 z + \sin \omega t), \sin(k_0 y + \cos \omega t)
\},
\label{GPflow}
\end{align} 
where $k_0=2\pi/L_0$, $\omega=2\pi/t_{\rm t}$, and $t_{\rm t}=L_0/u_0\approx 9\times 10^{-11}\,\rm s$ is the turnover time.  The flow is sustained by exerting a force of $\fv_i=C_f m_i \uv(\xv,t) /t_i$ on the ions, with the thermal ion passing time $t_i=L_0/\sqrt{2 T_i/m_i}$; we set $C_f=1$. The magnetic field is initialized as
$B_i=B_0 \sum_{j\ne i,n} b_{ij,n}\cos[n k  (x_i+\varphi_{ij,n})]$, where 
$b_{ij,n}$ and $\varphi_{ij,n}$ are uniform random numbers on $[0,1]$, $n=1,2,...,N$ with $N=4$, and $B_0=40$ (the thermal electron Larmor radius at this field strength is $2.7 \,\rm \mu m$). In addition to $\uv_{\rm GP}$, the initial electron flow velocity also has a component producing a current consistent with the magnetic seed field.

\begin{figure}
    \includegraphics[width=0.85\columnwidth]{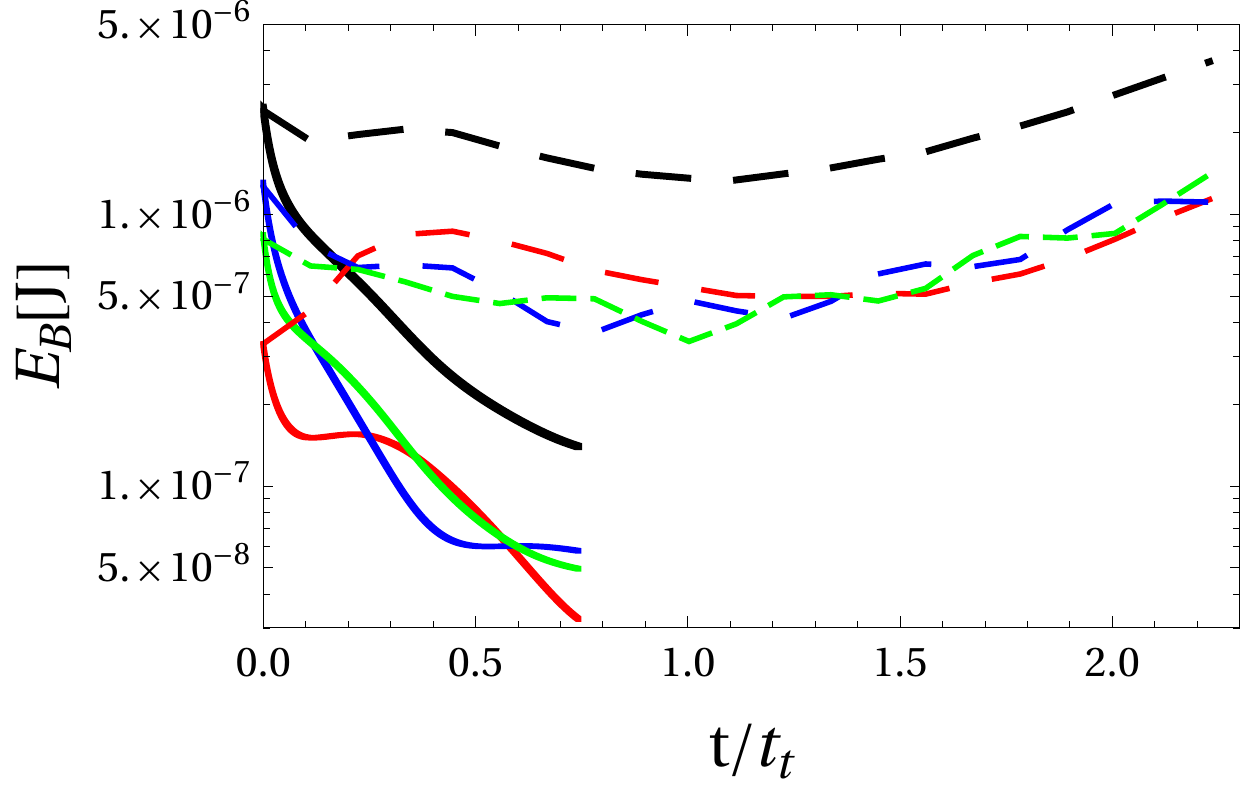}
   \put(-110,40){\normalsize kinetic}
   \put(-110,110){\normalsize MHD}
  \caption{\label{GPEnergies} Volume integrated magnetic energy. Solid lines:  kinetic simulation; dashed lines: resistive MHD induction equation. Red, blue, and green correspond to the contributions from $x$, $y$, and $z$ field components to the total (black). For reference, $(3/2) n_i T_i L_0^3=5.1\times 10^{-3}\,\rm J$.}
\end{figure}

\begin{figure}
    \includegraphics[width=0.8\columnwidth]{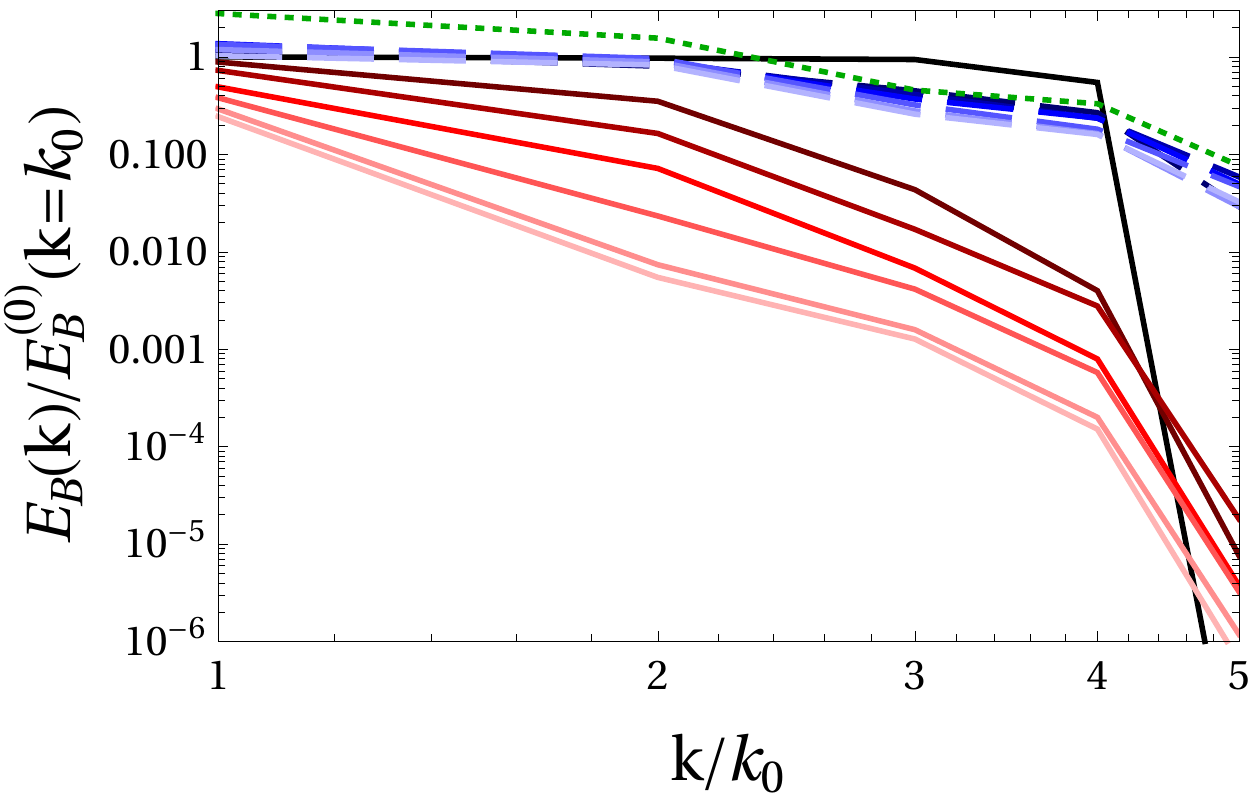}
  \caption{\label{kspec} Wave number spectra of magnetic energy, $E_B(k)$ (normalized to its value at $k_0$, $t=0$), for $t=\{0,1,2,...,6\}\times 10^{-11}\,\rm s$ (lines lightening). Solid lines: kinetic simulation; dashed lines: MHD induction equation; dotted line: MHD induction equation in the growing phase $t=2\times 10^{-10}\,\rm s$.  }
\end{figure}

The value $\rem\approx 13$ is sufficiently large for the GP flow to produce magnetic field growth in resistive MHD. Indeed, solving the MHD induction equation $\p_t \Bv=\nabla \times(\uv\times \Bv)+\eta \nabla^2 \Bv$ with $\uv=\uv_{\rm GP}$, using the high order finite-difference MHD solver {\tt Pencil~Code}  \cite{pencil} at spatial resolutions between $12^3$ and $32^3$, we find that, after a slight decay, the magnetic field starts to grow exponentially, as shown in Fig.~\ref{GPEnergies} (dashed). Additional MHD simulations (not shown here) also evolving the flow produce similar results. However, in the kinetic simulation, the field energy is observed to monotonically decay (solid). The magnetic energy in the kinetic simulation rapidly develops a strongly decaying wave number spectrum (solid lines in Fig.~\ref{kspec}). In contrast, the spectrum corresponding to the MHD induction equation quickly assumes its weakly decaying shape (dashed), which is then preserved in the phase of exponential growth (dotted line). The kinetic simulations used $12$ grid cells in each direction of the configuration space, $10$ in velocity space extending between $-3$ and $3$ times the thermal speed of each species, and employed a set of basis functions of polynomial order $1$, i.e., a resolution equivalent to $24$ and $20$ grid points, respectively, in a finite-difference scheme.  

The decay of the magnetic field energy in the kinetic simulation is caused by Landau damping of the magnetic fluctuations. To elaborate on this effect, we performed decaying magnetic field simulations in 1 spatial and 2 velocity coordinates, initialized with $B_z(x, t=0)=B_0 \cos (k x)$, and the corresponding current deposited as a flow of Maxwellian electrons in the $\mathbf{y}$ direction. The plasma parameters are similar to the GP flow simulation, and the simulations use up to 40 spatial and 20 velocity cells, with a polynomial order of 2.  For an elementary magnetic perturbation of this form, resistive magnetic diffusion $\p_t \Bv=\eta \nabla^2 \Bv$ leads to a decay $B_z\propto \exp(-\gamma t)=\exp(-k^2 \eta t)$. In a weakly collisional plasma, i.e., $\nu_{ei}\rightarrow +0$, where $\nu_{ei}$ is the electron-ion collision frequency, such a fluctuation decays due to Landau damping with a decay rate $\gamma =|k|^3c^2 v_e/(\sqrt{\pi}\omega_{pe}^2)=|k|^3 v_e m_e/(\sqrt{\pi}\mu_0 n_e e^2)$ \cite{mikhailovskii}, where $\omega_{pe}=\sqrt{n_e e^2/(\epsilon_0 m_e)}$, $v_e=\sqrt{2T_e/m_e}$ is the electron thermal speed, $-e$ and $n_e$ are the electron charge and density, and $\epsilon_0$ denotes the vacuum permittivity. We would get this decay rate from resistive diffusion, if we replaced $\sigma^{-1}$ with a scale-dependent effective resistivity $\sigma_{\rm eff}^{-1}=|k| v_e m_e/(\sqrt{\pi} n_e e^2)$, which corresponds to an effective magnetic diffusivity $\eta_{\rm eff}\sim \eta \lambda/l$, where $\lambda=v_e/\nu_{ei}$, and $2\pi/l=|k|$.
  
\begin{figure}
    \includegraphics[width=0.85\columnwidth]{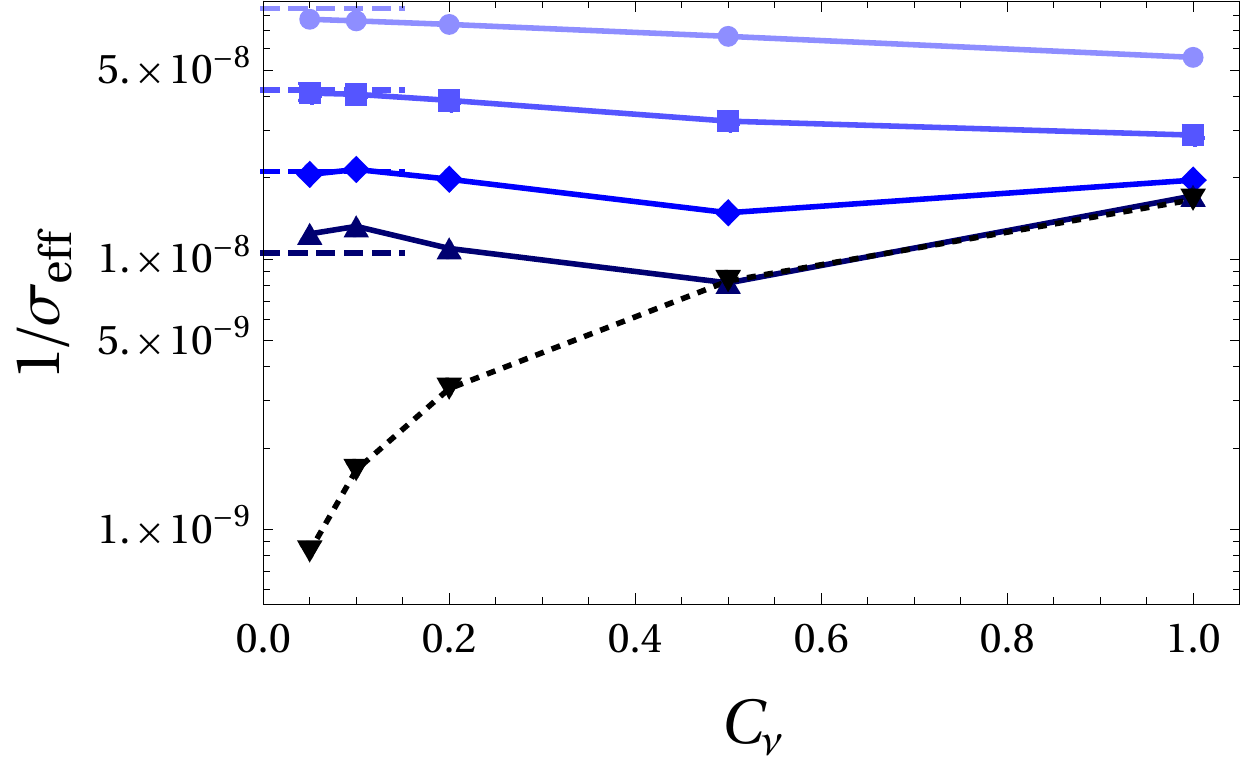}
   \put(-148,40){\small Spitzer}
   \put(-40,120){\small $L_0/8$}
   \put(-60,108){\small $L_0/4$}
   \put(-110,96){\small $L_0/2$}
   \put(-110,84){\small $L_0$}
  \caption{\label{sigmaeff} Effective resistivity, $1/\sigma_{\rm eff}\,  [\rm \Omega m]$, as a function of collision scaling factor $C_\nu$, for four values of the wavelength of the current perturbation, increasing from $L_0/8$ to $L_0=9.73 \,\rm \mu m$ (solid lines darkening). The Spitzer resistivity (dotted line), and the collisionless, unmagnetized theoretical limits (dashed lines) are also indicated.}
\end{figure}
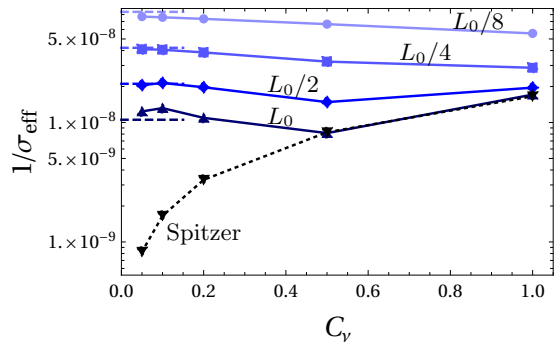

We introduce an overall collisionality scaling factor, $C_\nu$, that multiplies all inter- and intraspecies collision frequencies calculated for the given plasma parameters. Figigure.~\ref{sigmaeff} shows the $C_\nu$ dependence of $\sigma_{\rm eff}^{-1}$ that is calculated as an instantaneous value of $j_y/E_y$, and is consistent with the exponential decay rate of current perturbations. At the longest wavelength considered ($L_0=9.73 \,\rm \mu m$, dark solid curve) the effective resistivity starts deviating from the Spitzer resistivity below $C_\nu=0.5$, and for $C_\nu\rightarrow +0$ it asymptotes to a collisionality independent value determined by Landau damping. As the wavelength of the perturbations is decreased (lighter curves) the effective resistivity increases; in particular when $k=2 k_0$, $\sigma_{\rm eff}^{-1}$ remains  already above the Spitzer level over the collisionality range plotted. We note that perfectly collisionless simulations exhibit an echolike recurrence of the magnetic field energy, unlike the weakly collisional simulations shown here, where a simple exponential decay is observed.    

\begin{figure}
    \includegraphics[width=0.93\columnwidth]{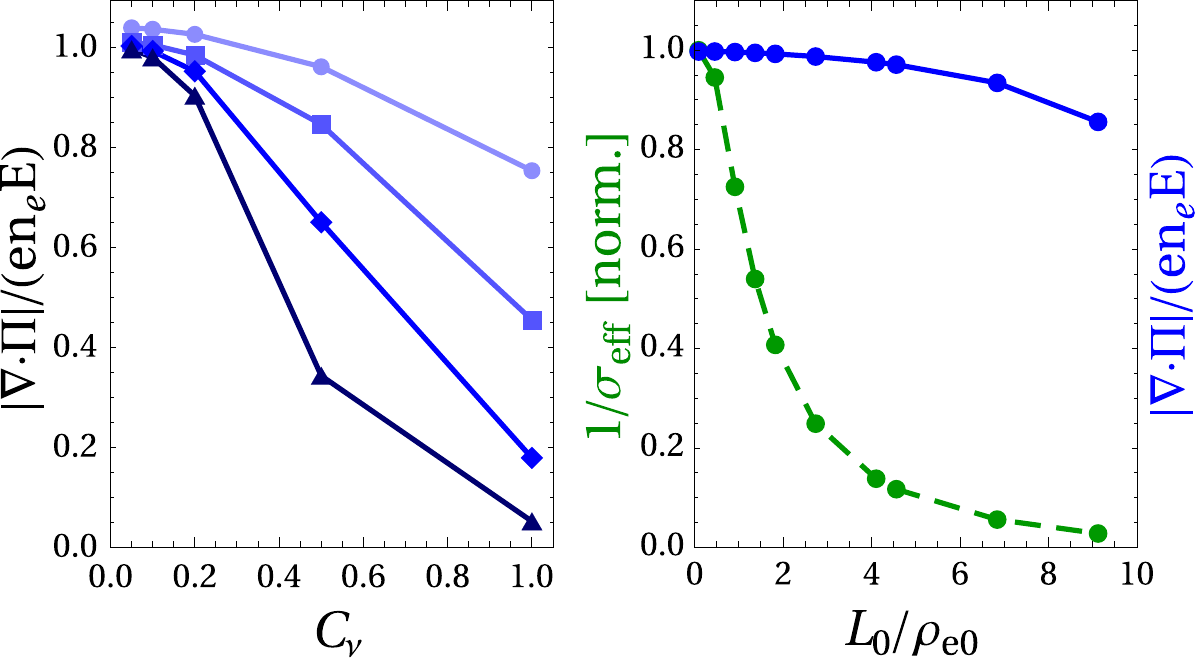}
   \put(-150,105){\small $L_0/8$}
   \put(-145,80){\small $L_0/4$}
   \put(-145,55){\small $L_0/2$}
   \put(-163,40){\small $L_0$}
   \put(-203,25){\small $\rm (a)$}
   \put(-90,25){\small $\rm (b)$}
  \caption{\label{forces} Solid lines: The ratio of the relevant component of the electron viscous stress and the electric field force. In  (a) the ratio is shown as a function of collision scaling factor $C_\nu$, for four values of the wavelength of the current perturbation, increasing from $L_0/8$ to $L_0=9.73 \,\rm \mu m$ (lines darkening). In (b) the ratio is shown as a function of the electron magnetization $L_0/\rho_{e0}$, where $\rho_{e0}$ is the electron Larmor radius at a field strength of $B_0$. Here, the effective resistivity is also shown (dashed curve, normalized to its highest value, $1.65 \times 10^{-8}\,\rm \Omega m$); the wavelength is $L_0$, and $C_\nu=0.05$.  }
\end{figure}
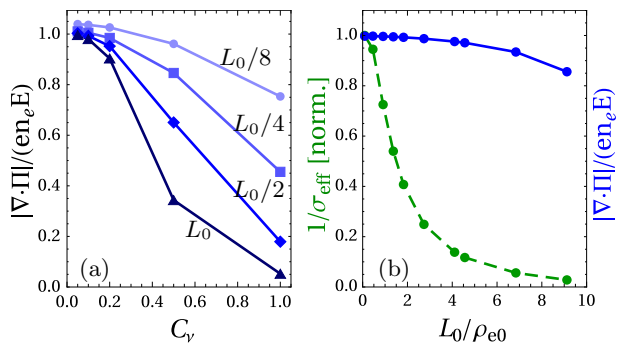

The simple physical picture behind the magnetic field decay in the collisionless regime is the following. A current perturbation of wave number $k$ would, without the self-consistent electromagnetic fields, decay on a time scale $\sim (v_e k)^{-1}$ due to free streaming; however, the corresponding $\partial_t \Bv$ induces an electric field that inhibits this current decay. The induced electric field being proportional to the current can be thought of as an effective resistivity, which leads to a diffusion, and thus a decay, of the magnetic field perturbation. In a collisional plasma, the electric field is balanced by collisional friction, resulting in a Spitzer response. In the weakly collisional case, however, the electric field is balanced by a viscous stress corresponding to an off-diagonal element of the electron pressure tensor, analogously to collisionless reconnection \citep{Vasyliunas,Birn2006,Hesse2011}. This viscous balance is illustrated in Fig.~\ref{forces}(a), where the ratio of the relevant viscous stress component to the electric force is shown as a function of $C_\nu$ for various wavelengths. In all cases, the small $C_\nu$ limit is close to unity, within a small difference due to electron inertia. The contribution from the viscous stress monotonically decreases with $C_\nu$ as the friction on ions becomes more important in balancing the electric field; at the longest wavelength (darkest curve) the viscous stress contribution is negligibly small for $C_\nu=1$, consistently with the Spitzer response observed in Fig.~\ref{sigmaeff}.    


Free streaming of electrons across the current perturbation is inhibited when the electrons are magnetized and are thus confined to magnetic field lines. Therefore, the Landau damping of magnetic field fluctuations becomes unimportant with increasing magnetic field strength, as illustrated in Fig.~\ref{forces}(b), showing the reduction of the effective resistivity with increasing $B_0$ ($\rho_{e0}$ is the electron thermal Larmor radius at $B_0$). For low $L_0/\rho_{e0}$, the $\sigma_{\rm eff}^{-1}$ is comparable to the theoretical collisionless value from Landau damping, and it drops rapidly with increasing $L_0/\rho_{e0}$. As for its relevance in dynamos, when the magnetic field energy grows, the range of scales where Landau damping of magnetic fluctuations are important decreases with the electron Larmor radius. 

\begin{figure}
    \includegraphics[width=0.78\columnwidth]{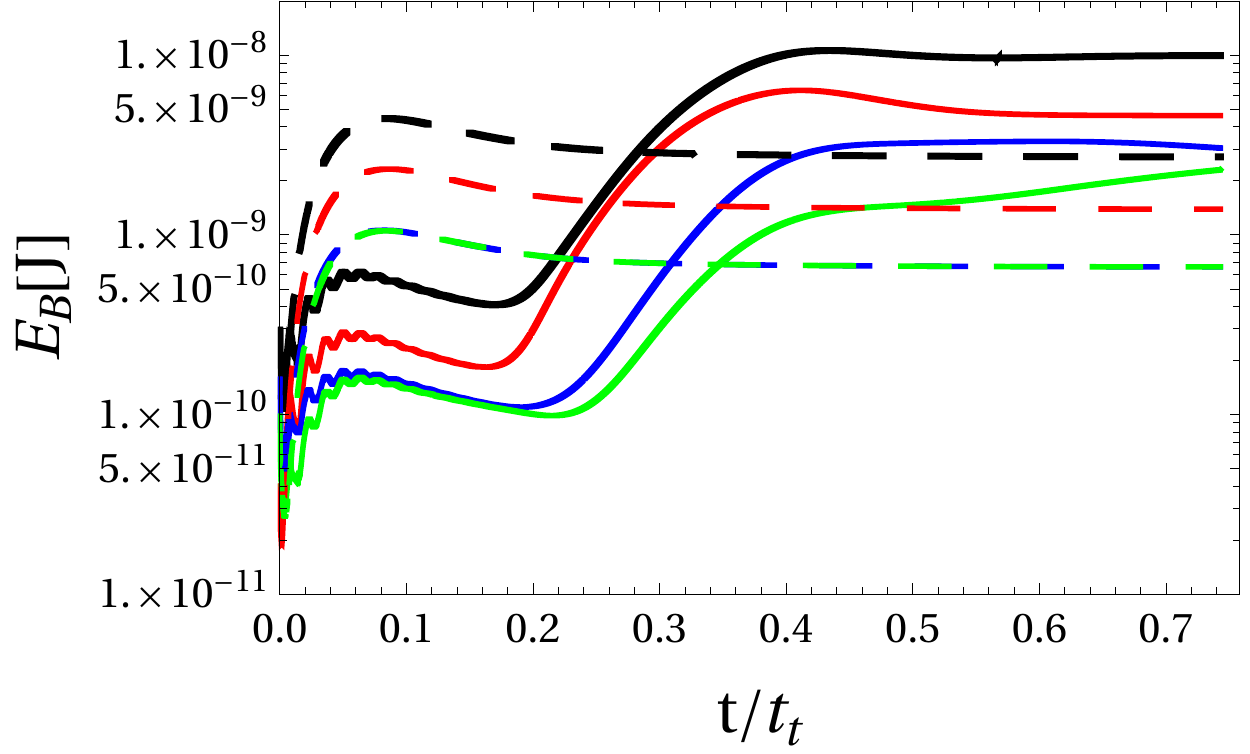}
  \caption{\label{robertsFig} Volume integrated magnetic energy in a forced Roberts flow simulation, for $C_\nu=0$ (solid) and $0.3$ (dashed). Red, blue, and green correspond to the contributions from $x$, $y$, and $z$ field components to the total (black).  
  For reference, $(3/2) n_i T_i L_0^3=9.94\times 10^{-6}\,\rm J$.}
\end{figure}

Note that accurate interpretation of fully kinetic dynamo simulations is made difficult by currents unavoidably driven by the forcing. Even exerting a force on ions and electrons appropriately scaled by their masses leads to a current, as the momentum transport properties of the two species are different (and magnetization-dependent); in weakly collisional plasmas, the corresponding driven current is comparable to that when forcing only acts on the ions. Therefore, a magnetic field is being generated that may be larger than the initial seed fields. This effect is illustrated in Fig.~\ref{robertsFig}, which shows the magnetic field energy in a simulation with a driven, time independent Roberts flow \cite{Roberts}
\begin{equation}
\uv_{\rm R} (\xv,t) = u_0 \{
\cos(k_0 y) - \cos(k_0 z),  \sin(k_0 z), \sin(k_0 y)
\}. 
\label{Rflow}
\end{equation}
In these simulations, $L_0=1.22 \,\rm \mu m$, $B_0=10\,\rm T$, the collisionality is scaled as $C_\nu=0$ (solid) and $C_\nu=0.3$ (dashed), and the flow is more strongly forced $C_f=3$, otherwise the parameters are similar to those of the Galloway-Proctor flow simulation. The magnetic field energies level off after an initial growth phase in both cases. We find that the final field strength is of the size $\sim e u_0 n_i \mu_0 L_0$, which is expected to arise from the forcing of the ion flow. Indeed, at the end of the simulations, the current density has a form close to $\uv_{\rm R}$ (as does $\Bv$, since the field is essentially force free). When simulations are started from a higher initial $B_0$, the magnetic energy decays down to the same level, where the continuous drive is balanced by the effect of Landau damping and collisions. For these parameters no dynamo amplification is observed in the simulation.       

Finally, we consider the implication of Landau damping on fluctuation dynamos with asymptotically large $\pra$. In the MHD framework, $l_\eta$ is estimated by balancing the rate of stretching of magnetic fluctuations at the viscous scale $u_\nu/l_\nu$ with the dissipation rate at the resistive scale $\eta/l_\eta^2$, yielding $l_\eta\sim l_0 \rey^{-3/4} \pra^{-1/2} \sim l_\nu \pra^{-1/2}$   \cite{Schekochihin_2004}. In a weakly collisional plasma, we may introduce the analogous Landau dissipation scale $l_L$, where magnetic field growth due to stretching at the viscous scale balances decay due to Landau damping. Thus, recalling $\eta_{\rm eff}(l)=\eta \lambda/l$, we balance $u_\nu/l_\nu\sim u_0/(l_0 \rey^{1/2})$ and $\eta_{\rm eff}(l_L)/l_L^2\sim \eta \lambda/l_L^{3}\sim \nu \lambda/(l_L^{3} \pra)$. This result, combined with $\lambda\sim l_0 M_0/\rey$, yields the estimate
\begin{equation}
l_L \sim l_0 \frac{M_0^{1/3}}{\rey^{5/6}\pra^{1/3}}\sim l_\nu \frac{M_0^{1/3}}{\rey^{1/12}\pra^{1/3}}.
\end{equation} 
When $\rey^{1/2} /M_0^2\ll \pra$, as for instance in galaxy clusters, $l_\eta \ll l_L$, implying that the range of scales over which magnetic field growth can occur is reduced compared to the prediction of resistive diffusion. 

In conclusion, considering weakly collisional, nonmagnetized initial conditions, we have performed fully kinetic continuum simulations of model flows known to produce dynamo amplification of the magnetic field in resistive MHD. The magnetic field energy---apart from that corresponding to a current caused by the forcing of the ion flow---in these cases is observed to decay due to the Landau damping of the magnetic perturbations. Demonstrating dynamo growth in this setting will demand an increased scale separation between the flows and the effective magnetic dissipation. The computational feasibility of greater scale separation would require employing reduced physics parameters, which we avoided here. The effect of the Landau damping is similar to that of a magnetic diffusivity that scales with the wave number of the perturbation $|k|$. In high magnetic Prandtl number plasmas (such as on galactic scales and above), the damping is expected to lead to a peak of the magnetic spectrum at $l_L$, a scale larger than that given by resistive diffusion, $l_\eta$, potentially reducing the total energy in magnetic fluctuations. As the magnetic field grows during the dynamo process, the scale at which electrons demagnetize decreases, shrinking the region where this process is operational. While the maximum of the saturated magnetic energy spectrum in kinetic ion hybrid simulations appears at the ion gyroradius scale \citep{StOnge}, our results suggest that a resolved and saturated fully kinetic dynamo simulation would produce a magnetic spectrum peaked around the electron gyroradius scale, or $l_L$, whichever is smaller. On scales where electrons are magnetized, the issue of magnetic moment conservation potentially impeding dynamo growth  \cite{helander_strumik_schekochihin_2016} becomes relevant. It is possible that, similarly to ions \cite{StOnge_phdthesis}, electrons develop their own instabilities and corresponding sharp phase space structures, leading to breaking magnetic moment conservation, and alleviating this problem. This remains to be demonstrated. 

The simulation data presented in this article is available at Zenodo \cite{zenodo}.

The authors are grateful for fruitful discussions with S.~L.~Newton, L.~Gremillet, S.~Tobias, F.~I.~Parra,  M.~Jenab, and T.~F\"{u}l\"{o}p.
This project has received funding from the European Research Council (ERC) under the European Union's Horizon 2020 research and innovation programme under Grant Agreement No 647121. J.M.T. was supported by the
NSF (SHINE award AGS-1622306), J.J.~by NASA (Earth and Space Science Fellowship, No.~80NSSC17K0428), and M.F.~by the U.S. Department of Energy (Grants No. DOE-SC-0010508 and No. DE-FC02-08ER54966). Simulations were performed on resources provided by Swedish National Infrastructure for Computing (SNIC) at HPC2N, and the Extreme Science and Engineering Discovery Environment, which is supported by National Science Foundation (No.~ACI-1548562).




\bibliography{kineticDynamo}

\end{document}